\documentstyle[11pt,newpasp,twoside]{article}
\begin{document}

\title{Star Cluster Simulations Including Stellar Evolution}
\author{Stephen L.~W.~McMillan}
\affil{Department of Physics, Drexel University, Philadelphia, PA 19104}

\begin{abstract}
The past few years have seen dramatic improvements in the scope and
realism of star cluster simulations.  Accurate treatments of stellar
evolution, coupled with robust descriptions of all phases of binary
evolution, have been incorporated self-consistently into several
dynamical codes, allowing for the first time detailed study of the
interplay between stellar dynamics and stellar physics.  The coupling
between evolution, dynamics, and the observational appearance of the
cluster is particularly strong in young systems and those containing
large numbers of primordial binary systems, and important inroads have
been made in these areas, particularly in $N$-body simulations.  I
discuss some technical aspects of the current generation of $N$-body
integrators, and describe some recent results obtained using these
codes.
\end{abstract}


\section{Introduction: A Revolution in Cluster Modeling}
Star clusters are remarkably clean realizations of the classical
$N$-body problem.  They are relatively isolated in space, consist of
more or less coeval stars, and contain relatively little gas or dust.
As such, they provide unique laboratories for the study of fundamental
stellar dynamical processes---relaxation, mass segregation, core
collapse, binary interactions, post-collapse evolution, and
evaporation with and without the effects of external tidal fields.
Globular clusters are perhaps the best examples of ``clean'' systems.
They are indeed isolated, gas free, and largely separate from the rest
of the Milky Way Galaxy (except for tidal effects), and stellar
evolution is relatively unimportant in governing their present
large-scale dynamics, although it may be critical in determining the
appearance of individual cluster components.

Of course, this is an oversimplified picture.  In reality, cluster
evolution involves not just stellar dynamics but also stellar
evolution and mass loss, binary evolution, stellar collisions and
tidal interactions, not to mention time-dependent external influences
such as molecular clouds, bulge and disk shocks, and the effects of
cluster orbital evolution.  Combined, this complex mixture of
competitive, tightly coupled physical processes poses great
difficulties to would-be modelers.  Open clusters and the Arches and
Quintuplet systems found in the Galactic nucleus are examples of such
``dirty'' systems (the term reflecting somewhat the prejudices of a
dynamicist).  Some are sites of ongoing star formation, gas dynamics
is evident in others, many are still strongly influenced by stellar
evolution, and almost all have relatively short lifetimes in the
Galactic tidal field.  Indeed, one could argue that the surviving
globulars are very unusual, and they are now clean only because they
have survived the competitive processes that have already destroyed 99
percent of their siblings.  Thus we need to study dirty systems even
to understand the clean ones observed today.

Studies of cluster dynamics came of age in the 1980s, when improved
simulation techniques elucidated many of the basic evolutionary
processes operating in stellar systems---core collapse, mass
segregation, binary dynamics, and the large-scale effects of mass loss
due to stellar evolution.  However, for good theoretical and practical
reasons, the work concentrated on problems consisting essentially of
point-mass dynamics with the inclusion of a few additional
well-defined effects of interest.  Limited combinations of physical
processes afforded better understanding of cluster dynamics, and
limitations in simulation techniques---the approximations inherent in
continuum codes and the sheer computational expense of $N$-body
simulations---largely dictated the simulation agenda.

The 1990s saw the integration of many detailed physical processes into
simulation codes.  Mass loss in evolving cluster environments, binary
dynamics in clusters, and eventually the introduction of both stellar
and binary evolution into $N$-body and Monte-Carlo codes.  At the same
time, the advent of special-purpose GRAPE hardware has greatly
increased the numbers of stars that can be followed by direct
integration.  The result is that realistic star-by-star simulations of
open clusters and small globular clusters are now possible, and the
prospects for expansion into larger globular clusters are good.  Our
simulations are still limited by our knowledge of detailed physics,
cluster initial conditions, gas dynamics, and orbital evolution, but
the state of the art today is clearly qualitatively different from
the situation even five years ago.

In this paper I outline some recent advances in cluster simulation,
with particular reference to the effects of stellar and binary
physics.  I first describe some basic dynamics and the effect of mass
loss due to stellar evolution.  I then turn to the simulation
techniques now in use, and specifically to their incorporation of
stellar and binary evolution.  Finally I list some recent studies
carried out using the new generation of $N$-body codes, and discuss a
few issues that remain to be resolved.


\section{Cluster Dynamics and the ``Kitchen Sink''}
\subsection{Dynamical Evolution}
An isolated cluster of mass $M$ consisting of non-evolving point stars
comes into dynamical equilibrium on a crossing time scale $t_D =
(GM/R_{vir}^3)^{-1/2}$, where $R_{vir}$ is the cluster virial radius.
It evolves thermally on a relaxation time scale $t_R \sim
(N/8\log\Lambda)\ t_D$, where $N$ is the number of stars in the system
and $\Lambda\sim0.1 N$ (Spitzer 1987, Giersz \& Heggie 1996).

Heat conduction from the cluster core to the outer halo precipitates
the phenomenon known as core collapse, in which the cluster's central
core shrinks to (formally) infinite density while the outer regions
expand (Antonov 1962).  The core-collapse time scale, for a cluster
consisting of equal-mass stars, is $t_{cc}\sim15 t_{R,h}$ (Cohn 1979),
where $t_{R,h}$ is the relaxation time at the half-mass radius
(comparable to the virial radius in most cases).  For our purposes,
the most important effect of core collapse is the creation of a dense
central core, leading to enhanced stellar and binary interactions.

Energy equipartition (dynamical friction) causes massive stars to
segregate to the cluster center (Spitzer 1969).  For a star of mass
$m$ in a cluster of mean mass $\langle m\rangle\ll m$, the segregation
time scale is $t_{seg}\sim t_R\,\langle m\rangle/m$.  With even a
modest range in masses, mass segregation can speed up core collapse by
a factor of 3--5 or more relative to the equal-mass figure (Inagaki
1985).  In systems with realistic mass spectra and broad ranges in
mass (0.1 to 100 solar masses is typical in very young star clusters),
significant mass segregation can occur in as little as $0.2 t_{R,h}$
(Portegies Zwart \& McMillan 2002a).  Since mass segregation leads to
the creation of a dense central core of high-mass stars, with the
attendant processes of binary formation and stellar interactions, mass
segregation effectively {\em is} core collapse in these cases.

Core collapse continues until some mechanism can replace the energy
lost by conduction from core to halo.  Leading contenders include mass
loss due to stellar evolution and collisions, which tend to unbind the
cluster, and heating due to binary interactions.  The dynamical effect
of binaries is most simply expressed by ``Heggie's Law'' (Heggie
1975)---the well-known statement that hard binaries (those having
energies greater than the mean stellar kinetic energy, $\frac32 kT$)
tend to become harder following interactions with other stars, while
soft binaries tend to be destroyed.  A population of primordial
binaries can support the core at a few percent of the half-mass radius
until the binaries are depleted by interactions or escape (Goodman \&
Hut 1989; McMillan \& Hut 1994).  Dynamical binary formation requires
core collapse (or collapse of a massive subsystem) down to a few
tens of stars.  In either case, binaries heat the system at a roughly
constant rate until they are destroyed by a collision or ejected from
the system.

Cluster dynamical evolution is also strongly influenced by external
(tidal) fields, which tend to strip the least bound, and often lowest
mass, stars.  Ultimately, tidal mass loss reverses core collapse by
removing the confining cluster potential, leading to the possibility
of a remnant core rich in binaries and/or massive stars if the cluster
lifetime is sufficiently short (McMillan \& Hut 1984; Portegies Zwart
et al.~2002a).  Typically, in a stationary tidal field corresponding
to a cluster moving in a circular orbit around the Galactic center,
the cluster mass goes roughly linearly to zero on a time scale of a
few times the ``tidal relaxation time'' (essentially the above
expression for $t_R$, but using the dynamical time at the tidal
radius), the details depending on the cluster's initial internal
structure (Portegies Zwart et al. 2001a).  Studies of time-dependent
fields suggest that the effects of external tidal shocks may well be
comparable to internal relaxation in determining cluster evolution
(Gnedin \& Ostriker 1997).

%
%
%

\subsection{Stellar Evolution}
Much useful information can be obtained by studying specific effects
in isolation and, given the choice, most workers would probably prefer
such an incremental approach.  However, this is not possible, as there
is little middle ground between models consisting of identical,
non-evolving point masses and the so-called ``kitchen sink''
simulations in which all relevant physical effects are modeled in
detail.\footnote{Paradoxically, the term refers to an expression
indicating completeness in which everything {\em but} the kitchen sink
is included.}

The reason for this is not difficult to grasp.  A cluster model
consisting of identical point masses is not an accurate description of
any real star cluster.  Such a simple system has only one relevant
parameter---the total number of stars, $N$.  However, increasing the
realism of the model by adding a spectrum of stellar masses
necessarily introduces specific stellar physics into the
calculation---a stellar mass function must be chosen, and the spatial
distribution of each stellar species defined.  Once this is done, it
immediately becomes necessary also to include the effects of stellar
evolution---stellar evolutionary time scales are often comparable to
the mass segregation/core collapse times of the parent cluster, so the
dynamical state of the cluster is tightly coupled to the state of the
component stars.  Alternatively, neglect of stellar evolution will
lead to a core populated with unphysically massive stars, whose
dynamics is unrepresentative of any real system.  In addition, real
star clusters contain large numbers of binary systems (e.g.~Rubenstein
1997) and stellar evolution drives binary evolution, so this too must
be incorporated if the simulation is to remain self-consistent.
Binary evolution may lead to stellar mergers and the production of a
veritable zoo of exotic objects (see Portegies Zwart et al.~2001b).

Thus, the seemingly innocuous improvement of including stellar masses
actually leads to a complex mix of physical processes and an
exceedingly difficult numerical problem.  As tidal effects and stellar
physics are included, it is unclear how well our insight extends from
simple to complex systems.  The interplay of physical processes
demands a more comprehensive approach.

\subsection{Dynamical Effects of Stellar Evolution}
The importance of stellar evolution to cluster dynamics was made clear
by Applegate (1986) in a semi-analytic study of stellar mass loss from
an idealized dynamical system.  The study revealed that, for
sufficiently flat initial mass functions (i.e.~enough mass in massive
stars), the combination of mass segregation and mass loss was
sufficient to unbind a cluster long before core collapse could occur.
For stellar masses distributed as a power law, $dN/dm \sim
m^{-\alpha}$ for $0.35M_\odot\la m \la 4 M_\odot$, Applegate found
that the dividing line between core collapse and disruption lay at
$\alpha\sim2$ (where the slope of the Salpeter mass function is
$\alpha = 2.35$).

These conclusions were refined and strengthened by Chernoff \&
Weinberg (1990), who carried out the first systematic survey of
globular cluster survival in Galactic environment with the effects of
stellar evolution included self-consistently into a dynamical model.
They employed a Fokker--Planck approximation with a simplified
``main-sequence to giant to remnant'' description of stellar evolution
and an approximate ``absorbing-boundary'' treatment of the Galactic
tidal field.  Their model clusters concentrated on initially
unsegregated King (1966) models with dimensionless depths $W_0 = 1,$
3, or 7 (corresponding to a range in cluster Galactocentric radii of
2--25 kpc for the cluster masses considered), with initial stellar
masses in the range $0.4M_\odot\le m \le 15 M_\odot$, distributed as
power-laws with $\alpha = 1.5,$ 2.5, or 3.5.  They found that all
clusters with $W_0=1$ or $\alpha = 1.5$ dissolved before core
collapse, while all clusters with $W_0=7$ and $\alpha > 1.5$
collapsed.  For $W_0=3$, the dividing line between disruption and
survival for a Hubble time lay at $\alpha$ somewhat greater than 2.5,
and those clusters that did survive lost most of their mass by the
present day.

The first systematic series of $N$-body calculations incorporating
stellar evolution was carried out by Terlevich (1987) using Aarseth's
NBODY5 integrator (Aarseth 1985).  A variety of power-law initial mass
functions was considered, and some models were started out of
dynamical equilibrium (an example of an initial condition inaccessible
to most other modeling techniques).  Computational limitations
restricted the systems studied to $\la1000$ stars, but the models
clearly demonstrated the same processes of mass segregation, mass
loss, and cluster dissolution seen in Applegate's semi-analytic
calculation and Chernoff \& Weinberg's Fokker--Planck simulations.
Model lifetimes were in good general agreement with observed open
clusters, and early termination of core collapse due to stellar
evolution was evident.

As available computing power increased, it became possible to compare
Fokker--Planck simulations directly with $N$-body calculations.  The
$N$-body studies by Fukushige and Heggie (1995) and Portegies Zwart et
al.~(1998) included more sophisticated treatments of stellar evolution
and external tides, although they were still limited to the low-mass
end of Chernoff \& Weinberg's study.  Fukushige and Heggie chose
initial model parameters (mass range, $\alpha$, and $W_0$) largely
identical to those of Chernoff \& Weinberg; the models of Portegies
Zwart et al.~used similar (King model) cluster initial profiles, but
adopted a more realistic initial mass function and a more
sophisticated treatment of stellar evolution.  Overall, both studies
found general qualitative agreement with the conclusions of Chernoff
\& Weinberg, although they disagreed in detail, particularly in cases
where cluster disruption occurred rapidly (and the Fokker--Planck
approximation is least applicable), when the $N$-body lifetimes
exceeded the Fokker--Planck estimates by as much as an order of
magnitude.

De la Fuente Marcos (1997) performed studies of the evolution of
small, tidally limited open clusters having a variety of initial mass
functions, with and without the inclusion of stellar (but not binary)
evolution.  All models started with a substantial fraction (1/3) of
primordial binaries having energies in the $\sim1$--$10 kT$ range.  He
found that the dissolution time scale of his models depended quite
sensitively on the choice of IMF, and that the binary population
shortly before dissolution could show characteristic features allowing
remnants of rich and poor clusters to be distinguished
observationally.  It is unclear how these results extend to larger
systems.


\section{Realistic Cluster Simulations}
Given that stellar mass loss is of crucial importance to cluster
evolution, and the basic ingredients have been known for over a
decade, one might reasonably wonder why it has taken so long for
stellar and binary evolution to be fully integrated into dynamical
simulations.  The short answer is that the methods best equipped to
handling the physics have until been recently been unable to follow
sufficiently large numbers of stars, as I now discuss.

\subsection{Simulation Methods}

The author's biases notwithstanding, a fairly convincing case can be
made that $N$-body codes are particularly well suited to the inclusion
of arbitrarily complex stellar (and other) physics.  They make no
simplifying assumptions about the geometry or physical state of the
system under study, include all interactions to all orders, and can in
principle be extended to incorporate any desired process.  The price
of this is computational expense.  At least for open clusters, and
possibly also for larger systems, it appears that direct $N^2$ force
methods are necessary for accurate long-term (post-collapse)
collisional calculations.

Tree codes (e.g.~Barnes \& Hut 1986) are a possible alternative with
superior [$O(N\log N)$] scaling, but they introduce spatial and
temporal correlations in interparticle forces, possibly causing
spurious diffusion effects which may be important during the
post-collapse phase.  McMillan \& Aarseth (1993) found that a tree
code could successfully reproduce the main features of cluster core
collapse, and specifically the collapse time scale, but their study
did not probe the post-collapse regime.  However, they found that
trees are inherently poorly suited to systems with individual particle
time steps and large dynamic ranges in space and time scales, leading
to significant complexity in the implementation of their tree code.
This latter consideration also applies to fast multipole methods
(Greengard \& Rokhlin 1987), which generally perform best in
homogeneous systems having limited dynamic ranges.

Direct-summation codes received an enormous boost in the 1990s with
the development of the GRAPE series of special-purpose computers
(Makino et al.~1997), the reason for this conference and the subject
of numerous contributions in these proceedings.  Abandoning
algorithmic sophistication in favor of raw parallel computing power,
the GRAPE family of processors increase the size of $N$-body systems
by perhaps an order of magnitude over what is feasible using
conventional supercomputer technology.  The advent of GRAPE has
greatly alleviated the computational bottleneck in $N$-body
simulations, making it possible to focus on the physics rather than
the numerics, and today GRAPEs are at the heart of all detailed
simulations of star clusters and dense stellar systems.  The current
state of the art in collisional $N$-body simulations is $N\sim10^5$.

Two popular competitors to direct summation are the use of direct
Fokker--Planck codes and gas-sphere methods.  In the former, the
effects of repeated stellar encounters are distilled into a set of
phase-space diffusion coefficients and the evolution of the system is
followed using the Fokker--Planck equation in phase space (Cohn 1979).
The latter methods transform the collisional Boltzmann equation into a
system of moment equations closely resembling those of stellar
structure, with stellar interactions providing the means for heat to
flow around the system (Bettwieser \& Sugimoto 1984).  Both approaches
are much cheaper than $N$-body techniques.  However, at least in their
simplest forms, they are not well suited to full treatments of cluster
evolution.  They rapidly become complex, unwieldy, and of questionable
statistical validity as new dimensions are introduced to accommodate
stellar masses, binary interactions, stellar collisions, etc.  In
addition, they admit only approximate treatments of tidal fields and
tidal stripping, although Takahashi and Portegies Zwart (1998) have
demonstrated that the tidal mass-loss rates can be largely reconciled
with detailed $N$-body results if proper care is exercised in applying
the stripping criterion.


Monte-Carlo methods provide, depending on one's point of view, an
alternative means of integrating the Fokker--Planck equation or a more
faithful realization of the probability distributions underlying it.
Either way, they represent promising ``middle ground'' between
$N$-body and continuum schemes, combining the ability to model large
systems with treatments of many physical effects.  Current codes
include both ``pure'' Monte-Carlo methods (Joshi, Nave, \& Rasio 2001,
Giersz 2001) and ``hybrid'' approaches combining Monte-Carlo
treatments of selected stellar species (binaries, compact remnants,
etc.) with continuous descriptions of the rest of the system (Giersz
\& Spurzem 2000).  As yet these schemes do not include detailed
treatments of stellar and binary evolution, but those developments are
anticipated within the next few years.

Thus---for now, at least---$N$-body methods seem to offer the best
means of including detailed stellar physics in cluster simulations,
despite their continuing limitations in system size.  Larger
simulations using Monte-Carlo methods are possible, and may include
more detailed physical modeling soon, but such methods are still not
well suited to systems characterized by large departures from
spherical symmetry or dynamical equilibrium.

\subsection{N-body Modeling}
The Aarseth ``NBODY'' series of $N$-body integrators (NBODY1--6) and
their offshoots (NBODY6++ and others) are widely known and quite
thoroughly documented (see Aarseth 1999 and Hurley, these
proceedings).  I will concentrate here on describing the ``Starlab''
software suite (Portegies Zwart et al. 2001b) and some of its
similarities and differences from the Aarseth codes.  While written as
an independent alternative to the NBODY series, Starlab shares many
structural and algorithmic features with it---perhaps not surprising,
since Sverre Aarseth has over the past three decades developed and
refined many of the key elements of all $N$-body simulations.
However, some significant differences exist too.

Starlab is a collection of modular software tools designed to simulate
the evolution of stars and stellar systems and to analyze the
resulting data.  The package consists of a library of loosely coupled
programs, sharing a common flexible data structure, which can be
combined in arbitrarily complex ways to study the dynamics of binary
and multiple star systems, star clusters and galactic nuclei.  Central
to Starlab is the {\tt kira} integrator, whose key features include
\begin{itemize}
	\item a fourth-order Hermite integrator (Makino \& Aarseth
		1992) using a block time step scheduling scheme
		(McMillan 1986)
	\item a tree-based data structure, with single stars and
		binary centers of mass forming the top-level nodes in
		the tree, replacing the simple arrays used in the
		NBODY codes
	\item use of GRAPE hardware for top-level nodes (when
		available)
	\item computation of all low-level (binary) motion relative to
		the center of mass 
	\item homogeneous treatment of binaries and multiples
		of arbitrary complexity using a binary tree structure;
		the use of relative coordinates and unperturbed motion
		where appropriate obviates the need for the  binary,
		triple and chain regularization schemes employed in
		the NBODY codes
	\item use of a self-consistent hierarchical unperturbed
		approximation for efficient treatment of isolated
		binaries and multiples
	\item an efficient treatment of lightly perturbed binaries,
		modeled on the ``slow KS'' formalism of Mikkola \&
		Aarseth (1996)
	\item stellar tidal interactions and collisions
	\item comprehensive treatments of stellar and binary evolution
\end{itemize}
The package is described in considerably more detail, and is available
for download, at {\tt http://manybody.org}.

As in NBODY4, stellar evolution is incorporated into {\tt kira} via
look-up from precomputed tracks, interpolating in age and mass as
described by Eggleton, Fitchett, \& Tout (1989), and in metallicity
following Hurley, Pols, \& Tout (2000; see also Hurley, these
proceedings).  Stars and binaries are evolved in time at regular
intervals or as needed, depending on individual circumstances
(e.g.~following a merger in a binary or a collision of unbound stars).

Binary evolution in {\tt kira} follows the work of Portegies Zwart
(1997), and includes internally consistent treatments of all relevant
evolutionary phases.  Detached binaries may experience tidal
circularization (Zahn 1978), slow mass loss due to stellar winds, more
rapid mass loss as a star becomes a white dwarf, or instantaneous mass
loss in a supernova explosion (Hills 1983).  The velocity ``kick''
that neutron stars, and possibly also black holes, receive at birth is
taken from Hartman (1997).  Angular momentum loss may occur through
stellar winds, magnetic braking (Rappaport, Verbunt, \& Joss 1983), or
gravitational radiation (Peters \& Mathews 1963).  For semi-detached
binaries, we include the possibility of tidal instability and a
comprehensive treatment of stable and unstable mass transfer.
Accretion disk formation around compact companions is also modeled.
The structure of the package is such that evolutionary recipes can be
updated and refined as theoretical insights improve, without
necessitating major modifications to the rest of the program.

Cross-sections for, and the most probable outcomes of, stellar
collisions not resulting from binary evolution are taken wherever
possible from published simulations, mainly using smoothed-particle
hydrodynamics (e.g.~Lombardi, Rasio, \& Shapiro 1996, Freitag \& Benz
2000, Sills et al.~2001).  Results in the literature tend to
concentrate on collisions of main-sequence stars and selected
encounters involving normal stars and specific compact objects of
interest ($0.5M_\odot$ white dwarfs, $1.4M_\odot$ neutron stars,
etc.).  For other collisions, we attempt to interpolate between
published results or choose the ``closest'' published calculation, or
we simply assign effective radii to the stars involved (so approach
within the sum of the effective radii implies merger).

These approximations are far from perfect, especially when one
recognizes that encounters in simulations usually occur in bound
systems, sometimes involving multiple stars, and thus rarely have
orbital parameters similar to those adopted in published results
(which typically assume parabolic orbits).  One can envisage more
sophisticated treatments, in which SPH or other simulations are
performed on the fly as needed and the resulting object is followed
back to thermal equilibrium using a stellar evolution code after
re-integration into the $N$-body system.  However, there are numerous
technical difficulties with this approach, not the least of them being
the problems of how to recognize automatically that an encounter is
``over,'' and how to characterize and evolve the merger product.

These issues are particularly critical in studies of runaway mergers,
where one massive star grows by consuming other stars in the core of a
young dense cluster (Portegies Zwart et al.~1999).  Here, detailed
collision cross sections, merger probabilities, the thermal evolution
of the merger product, and the evolution of very massive stars may all
play important roles in determining the final outcome.  For now, we
must simply try various parametrizations to assess the robustness of
the result, but it is hoped that these approximations can be improved
in future work.

Finally, we adopt a more or less agnostic attitude toward tidal
capture (Fabian, Pringle, \& Rees 1975; Press \& Teukolsky 1977), the
process of binary formation via tidal dissipation in an unbound close
encounter, whose effectiveness as a means of forming close binaries
remains controversial.  Potential applications include the formation
of LMXBs and CVs in the denser globular clusters (Verbunt \& Hut
1987).  However, the process requires orbital separations comparable
to the radii of the stars involved, so the possibility of collisions
between the (probably thermally expanded) stars places a lower limit
on the allowed separation (McMillan, McDermott, \& Taam 1987; Ray,
Kembhavi, \& Antia 1987).  An upper limit is imposed by the
possibility that the binary becomes unbound on subsequent periastron
passages (Kochanek 1992).  The circularization process is discussed in
detail by Mardling (1995) and Mardling \& Aarseth (2001), and their
results are incorporated into NBODY4.  In {\tt kira}, the unknown
stellar state following the first encounter is viewed as the primary
factor limiting our knowledge of the process, and we generally simply
increase the effective stellar collision radius, although we retain
the option (usually switched off) of a more detailed treatment should
models improve.

\subsection{Some Applications}

Rather than describing a few specific applications of the modeling
techniques just described, I simply list here some recent examples and
refer the reader to more detailed accounts elsewhere in these
proceedings and in the literature.

Studies of open clusters range from simulations of the dynamics and
appearance of small systems such as the Pleiades, Hyades, and Praesepe
(Portegies Zwart et al.~2001b, 2002b; Starlab) to a detailed analysis
of binaries and blue stragglers in M67 (Hurley, these proceedings;
Hurley et al.~2001; NBODY4).  Shara \& Hurley (2002; see also Shara,
these proceedings; NBODY4) have found that dynamical interactions in
clusters containing primordial binaries can lead to a substantial
population of white-dwarf binaries---greatly exceeding expectations
based on binary evolution alone---with inspiral time scales less than
a Hubble time.  If these results can be scaled to globular clusters,
they may imply an interesting new channel for the production of Type
Ia supernovae.

Portegies Zwart et al (1999; Starlab) performed simulations of young
clusters, modeling the evolution of R136 in the 30 Doradus region of
the LMC.  They found that, for sufficiently dense (but not
unreasonably so) initial conditions, massive stars can segregate to
the cluster center and undergo runaway merger, leading to the
formation of stars having masses hundreds of times that of the Sun in
less then a few million years.  More recently, Johnson et al.~(2001)
have compared population-synthesis and dynamical simulations of the
young star clusters NGC 1805 and 1818 in the LMC with HST observations
of these systems.

Portegies Zwart et al.~(2001a, 2002a; see also Portegies Zwart, these
proceedings; Starlab) have performed a series of simulations of the
lifetimes and visibility of Arches- and Quintuplet-like clusters near
the Galactic center, concluding that the central regions of our Galaxy
may harbor dozens of similar clusters below the threshold of
visibility.  Subsequent studies by Portegies Zwart \& McMillan (2002)
and McMillan \& Portegies Zwart (2002) have followed both the inspiral
of such clusters into the Galactic Center and the possible formation
of intermediate-mass black holes following runaway mergers in their
cores.  Ebisuzaki et al.~(2002; see also Ebisuzaki, these proceedings)
have extended these ideas to the formation and subsequent inspiral of
intermediate-mass black holes in M82 and other galaxies.


\section{The Future}

The state of the art in cluster simulations has made impressive
strides, but many outstanding issues remain.  I conclude by listing,
in no particular order, some significant challenges to be addressed in
the next few years.

\subsection{Cluster Initial Conditions}
All simulations necessarily make simplifying assumptions about the
initial state of the cluster under study.  They start with a specific
choice of IMF (power-law, Scalo, Kroupa, etc.) usually based on
observations of stars in the solar neighborhood.  All stars are
assumed to lie on the zero-age main sequence, the cluster is taken to
be gas-free and in dynamical equilibrium (but see Kroupa, Aarseth, \&
Hurley 2001), the density profile takes on some simple standard form
(e.g.~Plummer, King, anisotropic King) and generally is applied to
stars of all masses.  Binary parameters are chosen from convenient
(but poorly known) distributions, with binary components typically
drawn randomly from the IMF and binary spatial densities following the
master density profile of cluster stars.
\par\noindent
Of course, most of these assumptions are incorrect.
\begin{itemize}

\item There is no reason to suppose that the local stellar mass
function or binary population is representative of initial mass
functions in young globular clusters (billions of years ago), or in
dense massive systems forming today (see Kroupa 2001 for a thorough
discussion).

\item Stars of different masses do not all reach the zero-age main
sequence simultaneously, so the births of the ``coeval'' stars in our
models may easily span 100 Myr or more, longer than the cluster
lifetime in some cases; protostellar disks may have large cross
sections and might contribute significantly to collisional mergers.

\item Many, if not most, young star clusters are rich in gas whose
eventual expulsion will cause the cluster to expand, and perhaps
dissolve.  Thus, clusters are probably not in dynamical equilibrium at
birth (or when stellar dynamics is typically ``turned on''), and gas
dynamics should be routinely incorporated into models of early cluster
evolution.

\item It is also possible that clusters are born mass segregated, with
more massive stars forming in the densest regions of the parent cloud
(see the discussions by Klessen and Whitworth, these proceedings), so
the assumption of a universal density profile may not be warranted.
In that case, the use of mass segregation as an indicator of dynamical
evolution must also be reconsidered.
\end{itemize}

\subsection{Better Treatments of Stellar Physics}
As described above, look-up tables, semi-analytical recipes, and
specific simulations now form the basis for the inclusion of stellar
and binary evolution and collisions into cluster simulations.  Can we
do better?  Might actual stellar evolution, SPH, even binary evolution
codes be embedded in $N$-body and other integrators?  Technically,
such an interface would be quite feasible---for example, the internal
structure of {\tt kira} was designed with the inclusion of both
stellar evolution and SPH in mind---although a clear formulation of
the possible interactions between the various modules involved is
critical.

The basic challenge of such an approach is to build a robust
standalone module that can function reliably and reasonably no matter
how unexpected or unreasonable the input might be.  While numerous
capable stellar evolution packages exist, none is intended to follow
the evolution of an object from birth to death without user
intervention at certain key stages.  Nevertheless, such a
self-contained program could probably be constructed, given a
sufficiently motivated programmer.  Experiments with Starlab indicate
that SPH can also be incorporated, although this possibility has never
been pursued to a final working product.  Binary evolution appears
much harder, as the physics is less well understood, and the prospect
of robust automatic evolution seems remote.

If any of this turns out to be feasible, we should then ask if it is
desirable.  The obvious advantage is that we can follow in detail the
evolution of any object, be it an original star or the result of a
later merger, without the need for look-up tables or approximate
``rejuvenation'' schemes for locating a merger product in an existing
array of precomputed models.  On the other hand, the automatic module,
in part by the very nature of the robustness needed for standalone
use, may result in models of lower resolution than those underlying
the look-up tables.  In addition, one could argue that the ability to
parametrize physical processes into recipes allows us the freedom to
experiment with individual effects (much as Monte-Carlo methods do
with dynamics) that perhaps could not be isolated within a more
complete evolution treatment.  Most probably the coming 2--3 years
will see the development of hybrid methods combining look-up and
automated computation of stellar structure as needed.

\subsection{Scaling and Calibration}
Even with GRAPE-6, it is still not feasible to model large globular
clusters using $N$-body methods.  Can we scale up smaller runs to tell
us about larger systems?  In the early 1990s, the hope was that
studies of cluster evolution of steadily increasing size would lead to
the development of simple scaling relations that would permit
extrapolation of $N$-body results to larger values of $N$.
Unfortunately, it now seems that there are too many ``important'' time
scales for that to be possible: in addition to the dynamical and
relaxation time scales (\S2.1), cluster evolution also depends on the
evolutionary time scales $t_S$ for individual stars, the time scales
for collisions between stars, and the time scale on which stars are
stripped by an external field.  All of these scales depend on the
choice of stellar mass function and other cluster parameters.

The ratio $t_R/t_D$ is set by the value of $N$.  The basic approach
used by researchers since Chernoff \& Weinberg (1990) has been to try
to preserve one of the ratios $t_S/t_D$ or $t_S/t_R$.  However,
Portegies Zwart et al.~(1998) found that the scaling even from 4k to
16k is unreliable in presence of tides and stellar mass loss,
whichever of these two ratios is held fixed.  Variable scaling
(Aarseth \& Heggie 1998), in which one adjusts the conversion between
stellar and dynamical time scales as the evolution proceeds and the
system makes a transition from ``dynamical'' to ``relaxation''
scaling, may work in certain controlled situations, but it appears
inadequate in the presence of primordial binaries, binary evolution,
and the possibility of stellar collisions.  The introduction of
time-dependent external fields only makes matters worse.  With scaling
efforts currently in doubt, there seems no alternative to doing it the
hard way, in which case realistic $N$-body models of million-body
systems may have to await the arrival of the GRAPE-8.

Other methods, especially Monte-Carlo codes, have already reached the
million-body mark (see Giersz 2001), and we can expect them to
incorporate realistic stellar physics in due course, as already
discussed.  However, in light of the simplifying assumptions made by
these methods, $N$-body calibration will be critical as the scope of
the effort expands.  With more than one independent version of each
major type of code now available, direct comparison of different
simulation methods (e.g.~Heggie, these proceedings) becomes both
feasible and essential.

\subsection{Modeling Individual Systems}
Finally, it is likely that the coming years will see increasing
numbers of end-to-end simulations of specific systems, with the goal
of making direct comparison with observations.  Examples of such
studies can be found in the recent simulations reported by Hurley et
al.~(2001), Portegies Zwart et al. (1999, 2001ab), and Johnson et
al.~(2001).  While $N$-body studies of the larger globular clusters
remain some years away, a number of intermediate-scale scale studies
of systems containing $\sim1$--$2\times10^5$ stars are now in
progress, with stellar physics and realistic (i.e.~non-circular)
cluster orbits taken properly into account.  As the realism and
resolution of simulations increase, visualization techniques such as
those described by by Levy (these proceedings) will grow in
importance, as will the use of standardized data-archiving tools to
analyze and distribute simulation results (Teuben, these proceedings).

\acknowledgements This work was supported in part by grants NAG5-9264
and NAG5-10775 from the NASA Astrophysics Theory Program, and by the
Sloan Foundation.  I also gratefully acknowledge the hospitality of
Tokyo University and the American Museum of Natural History in New
York.

\end{document}